\documentclass[aps,pra,twocolumn,footinbib]{revtex4-1}
\usepackage{graphicx}
\usepackage{indentfirst}
\usepackage{braket}
\usepackage{float}
\usepackage{amsmath}
\usepackage{epstopdf}
\usepackage{CJK}
\usepackage{esint}
\usepackage{color}
\usepackage[T1]{fontenc}
\usepackage{subfigure}
\usepackage{amsfonts}
\usepackage{footmisc}
\usepackage{scrextend}
\usepackage{multirow}
\usepackage{cleveref}
\usepackage[english]{babel}
\usepackage{url}

\newcommand{\bp}{\boldsymbol \rho}
\newcommand{\bn}{\boldsymbol \eta}

\def\be{\begin{equation}}
\def\ee{\end{equation}}
\def\ba{\begin{eqnarray}}
\def\ea{\end{eqnarray}}
\usepackage{bm}

\begin{document}

\title{Distributed quantum sensing using continuous-variable multipartite entanglement}

\author{Quntao Zhuang$^{1,2}$}
\email{quntao@mit.edu}
\author{Zheshen Zhang$^{1,3}$}
\author{Jeffrey H. Shapiro$^1$}
\affiliation{$^1$Research Laboratory of Electronics, Massachusetts Institute of Technology, Cambridge, Massachusetts 02139, USA\\
$^2$Department of Physics, Massachusetts Institute of Technology, Cambridge, Massachusetts 02139, USA\\
$^3$Department of Materials Science and Engineering, University of Arizona, Tucson, Arizona 85721, USA}
\date{\today}

\begin{abstract} 
Distributed quantum sensing uses quantum correlations between multiple sensors to enhance the measurement of unknown parameters beyond the limits of unentangled systems. We describe a sensing scheme that uses continuous-variable multipartite entanglement to enhance distributed sensing of field-quadrature displacement.  By dividing a squeezed-vacuum state between multiple homodyne-sensor nodes using a lossless beam-splitter array, we obtain a root-mean-square (rms) estimation error that scales inversely with the number of nodes (Heisenberg scaling), whereas the rms error of a distributed sensor that does not exploit entanglement is inversely proportional to the square root of number of nodes (standard quantum limit scaling). Our sensor's scaling advantage is destroyed by loss, but it nevertheless retains an rms-error advantage in settings in which there is moderate loss.  Our distributed sensing scheme can be used to calibrate continuous-variable quantum key distribution networks, to perform multiple-sensor cold-atom temperature measurements, and to do distributed interferometric phase sensing.

\end{abstract} 

\maketitle
\section{Introduction}
Single-mode squeezed states enable metrology beyond the standard quantum limit (SQL).  In particular,  they can increase the sensitivity of the Laser Interferometer Gravitational-Wave Observatory (LIGO)~\cite{LIGO_nat,LIGO}, and enable sub-shot-noise biological imaging~\cite{taylor2013biological}. Entanglement, on the other hand, possesses nonlocal properties that single-mode squeezing does not offer. For example, when the bipartite entanglement of two-mode squeezed states is leveraged in target detection, it provides a signal-to-noise ratio advantage over that of the optimum classical scheme~\cite{Lloyd2008,Tan2008,Guha2009,Zheshen_15,Zhuang2017}. Prior work has shown that multipartite entanglement between distributed sensors could yield significant sensitivity enhancement in estimating the weighted sum of unknown parameters in the sensor network~\cite{ge2017distributed,proctor2017multi}. However, these distributed quantum-sensing protocols rely on photonic discrete-variable (DV) multipartite entanglement, which, to date, can only be probabilistically generated and is extremely vulnerable to environmental loss. Such scalability disadvantage hinders DV distributed quantum-sensing protocols' being applied in practical situations.

Continuous-variable multipartite (CVMP) entanglement, in contrast, is highly scalable, because it can be deterministically generated, distributed, and detected~\cite{yukawa2008experimental}. What is equally important is that the quality of CVMP entanglement degrades gracefully in the presence of loss. As such, CVMP entanglement opens an attractive path toward scalable and distributed quantum sensing with robustness to loss. In this paper, we derive the optimum CVMP entangled state for distributed sensing of field-quadrature displacement, and find that the optimum state, produced by dividing a single-mode squeezed-vacuum state with a lossless beam-splitter array, achieves Heisenberg-scaling sensitivity~\cite{zwierz2010general,Giovannetti_2001,giovannetti2004,giovannetti2006,giovannetti2011advances,escher2011,refereeB1} in the number of sensing nodes in a network.  Moreover, although the entangled state's performance loses its Heisenberg scaling in the presence of loss, it retains a performance advantage over individually operating sensing nodes in moderate loss.  Furthermore, its implementation only requires the available technologies of squeezed-vacuum generation, linear optics, and homodyne detection.   

The emergence of quantum networks~\cite{Qinternet}, e.g., with fiber-optic connections in metropolitan areas~\cite{sasaki2011field}, or with satellite-communication connections~\cite{wang2013direct} over longer distances, offers a variety of application scenarios for distributed sensing.  Many continuous-variable 
quantum key distribution (CV-QKD) protocols rely on field-quadrature displacements~\cite{grosshans2002continuous,pirandola2006continuous,pirandola2015high}, and our sensing scheme could improve joint calibration of systematic errors in displacement operations in such network settings.  Ultrahigh-precision interferometric phase sensing can be reduced to field-quadrature displacement measurement for which quantum enhancement can be valuable.  Indeed, as analyzed in Refs.~\cite{demkowicz2013fundamental,escher2011,schnabel2010quantum} for a model of LIGO and experimentally demonstrated in Ref.~\cite{LIGO_nat}, single-mode squeezed-vacuum injection improves the performance of a single interferometer.  For multiple, spatially-separated, interferometers, our distributed displacement sensor can offer a further quantum enhancement by replacing each interferometer's single-mode squeezed-vacuum input with its portion of a CVMP entangled state.  Additional, more localized, applications of our distributed field-quadrature sensor arise in cold atom systems.  There, angular momentum~\cite{eckert2008quantum} and temperature measurements~\cite{mehboudi2015thermometry} can be reduced to field-quadrature displacement measurements, allowing our approach to afford increased sensitivity in multi-node sensing configurations

Before proceeding, it is worth contrasting the approach we will take with recent work on distributed quantum sensing~\cite{ge2017distributed,proctor2017multi}.  Ref.~\cite{ge2017distributed}'s distributed phase sensing required twin Fock-state generation and photon-number resolving detectors to realize Heisenberg scaling, and Ref.~\cite{proctor2017multi}'s contribution was a general framework for distributed sensing showing that measurement precision in estimating the weighted sum of unknown parameters in the sensor network could be improved by employing the multipartite entanglement of Greenberger-Horne-Zeilinger states. Our paper presents an explicit distributed-sensor design whose CVMP entanglement generation and distributed quantum measurement can easily be realized.

\section{Distributed field-quadrature sensing} Consider a network of $M$ sensing nodes each of whose optical input (with annihilation operator $\hat{a}_m$, for $1\le m \le M$) undergoes an identical real-valued quadrature displacement $\alpha$ by the unitary transformation $\hat{U}(\alpha)$. Our goal is to find the joint state, $\hat{\bp}_{M,N_S}$, for the $\{\hat{a}_m\}$ that: (1) contains $N_S$ photons on average; and (2), after distribution to the sensor nodes through pure-loss channels with transmissivity $\eta$, minimizes the root-mean-square (rms) error in estimating $\alpha$ from the ideal-homodyne quadrature measurements, $\{{\rm Re}(\hat{a}'_m) \equiv \sqrt{\eta}\,{\rm Re}(\hat{a}_m) + \alpha + \sqrt{1-\eta}\,{\rm Re}(\hat{e}_m): 1\le m\le M\}$ with the $\{\hat{e}_m\}$ being vacuum-state modes~\cite{footnote3}. (Appendix~\ref{AppA} shows that our protocol can be adapted for advantageous sensing of the weighted sum of different displacements at each node when the transmissivities to each node are also different but known.)

In the remainder of this section we will derive the optimum entangled and separable $\hat{\bp}_{M,N_S}$ for this distributed sensing problem, and compare their rms estimation errors.  

\begin{figure}
\centering
\includegraphics[width=0.35\textwidth]{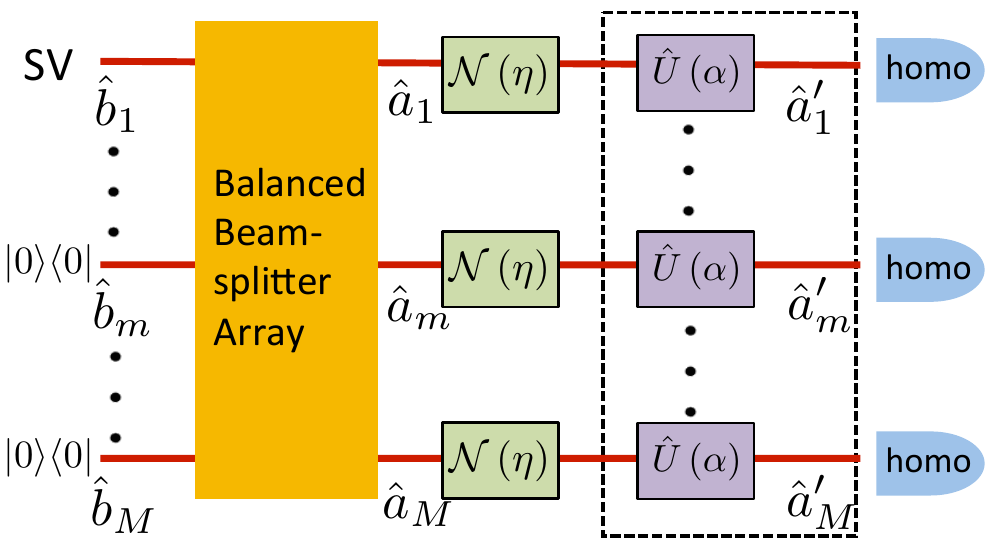}
\caption{Distributed quantum sensor for measuring field-quadrature displacement.  
SV: squeezed-vacuum state with mean photon number $N_S$ and squeezed noise in its real quadrature. $\mathcal{N}(\eta)$:  pure-loss channel with transmissivity $0< \eta \le 1$.  $\hat{U}(\alpha)$:  field-quadrature displacement by real-valued $\alpha$.  homo: homodyne measurement of the real quadrature.   
\label{int1}
}
\end{figure}

\subsection{Optimum Entangled State}
The joint state at the inputs to the sensors nodes' homodyne detectors is
$
\hat{\bp}_{M,N_S,\eta}\left(\alpha\right)=\hat{U}(\alpha)^{\otimes M} [\mathcal{N}(\eta)^{\otimes M} (\hat{\bp}_{M,N_S})]\hat{U}(\alpha)^{\otimes M\dagger},
$ 
where $\mathcal{N}(\eta)$ denotes a pure-loss channel with transmissivity $0< \eta \le 1$.  The displacement $\alpha$ only contributes to the the homodyne measurements' mean values, $\{\langle {\rm Re}(\hat{a}'_m)\rangle = \alpha + \sqrt{\eta}\,\langle {\rm Re}(\hat{a}_m)\rangle\}$, so we shall use 
\be
\tilde{\alpha}_E \equiv \frac{1}{M}\sum_{m=1}^M[{\rm Re}(\hat{a}'_m-\sqrt{\eta}\,\langle \hat{a}_m\rangle)],  
\ee
as our displacement estimator.
By introducing $\hat{b}_1 \equiv \sum_{m=1}^M\hat{a}_m/\sqrt{M}$, we can rewrite $\tilde{\alpha}_E$ as 
$\tilde{\alpha}_E = {\rm Re}(\hat{b}'_1 - \sqrt{\eta}\,\langle \hat{b}_1\rangle)/\sqrt{M}$,
where $\hat{b}'_1 \equiv \sqrt{\eta}\,\hat{b}_1 + \sqrt{M}\,\alpha + \sqrt{1-\eta}\,\hat{e}$, with the $\hat{e}$ mode being in its vacuum state.  It immediately follows that $\tilde{\alpha}_E$ is an unbiased estimator, $\langle \tilde{\alpha}_E\rangle = \alpha$, whose rms estimation error is
\be
\delta\alpha^E_\eta = \sqrt{[\eta{\rm Var}[{\rm Re}(\hat{b}_1)] + (1-\eta)/4]/M},
\ee
where ${\rm Var}(\cdot)$ denotes variance.  So, to make optimum use of the light available under the $\{\hat{a}_m\}$'s average photon-number constraint, we will assume that these modes are obtained from passing modes $\{\hat{b}_m: 1\le m \le M\}$ through a lossless, $M \times M$ balanced beam splitter with the $\hat{b}_1$ mode having average photon number $N_S$ while the other $M-1$ inputs, $\{\hat{b}_m : 2\le m \le M\}$, are in their vacuum states.  With this beam-splitter arrangement, each $\hat{a}'_m$ mode is comprised of a $\sqrt{\eta}\,\hat{b}_1/\sqrt{M}$ component plus vacuum contributions from the $\{\hat{b}_m : 2 \le m \le M\}$ and $\hat{e}_m$ modes and the quadrature displacement by $\alpha$, as shown in Fig.~\ref{int1}.  Well known properties of single-mode squeezed states~\cite{Yuen_Shapiro_1978} then imply that $\delta\alpha^E_\eta$ is minimized if the $\hat{b}_1$ mode is in its squeezed-vacuum state with average photon number $N_S$ whose real quadrature is squeezed.  The resulting rms error for this optimum entangled-state input is 
\be
\delta \alpha_\eta^E=\frac{1}{2}\left(\frac{\eta}{M\left(\sqrt{N_S+1}+\sqrt{N_S}\right)^2}+\frac{1-\eta}{M}\right)^{1/2}.
\label{dalpha_E}
\ee

The preceding performance exhibits Heisenberg scaling, in the lossless case, with respect to the number of sensor nodes.  Specifically, when $\eta =1$ and the average photon number per node, $n_S \equiv N_S/M\gg 1$, is kept fixed, we have that $\delta\alpha_1^E \simeq 1/4M\sqrt{n_S}$, whereas for the optimum separable-state $\hat{\bp}_{M,N_S}$, which we derive below, the rms error when $\eta = 1$ and $n_s \gg 1$ is fixed has SQL scaling, viz., $\delta\alpha_1^P \simeq 1/4\sqrt{Mn_s}$.  We postpone further discussion of Eq.~(\ref{dalpha_E}) until after we obtain the optimum separable-state $\hat{\bp}_{M,N_S}$ for our distributed-sensing problem.

\begin{figure*}
\subfigure{
\centering
\includegraphics[width=0.225\textwidth]{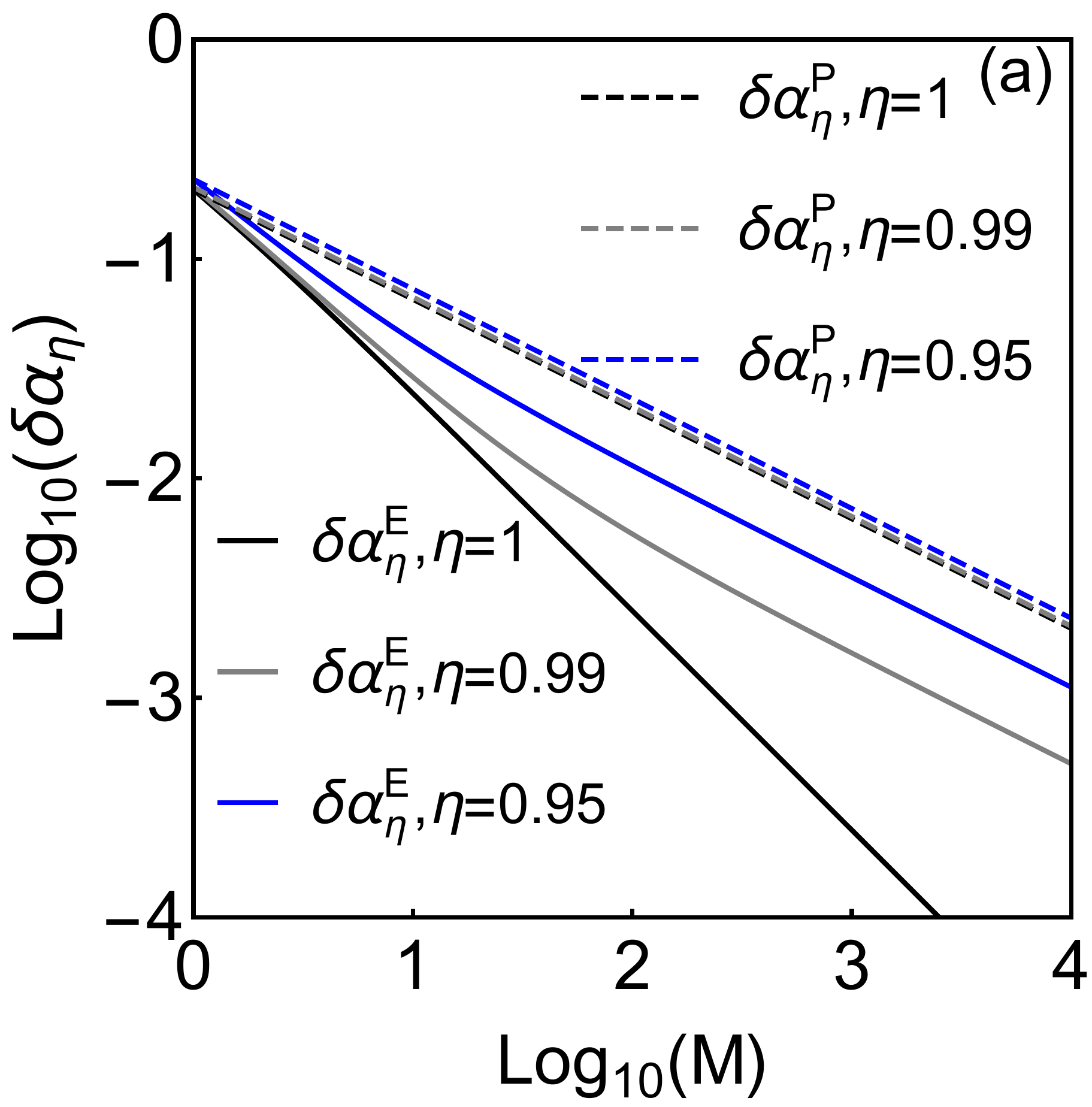}
} 
\subfigure{
\centering
\includegraphics[width=0.225\textwidth]{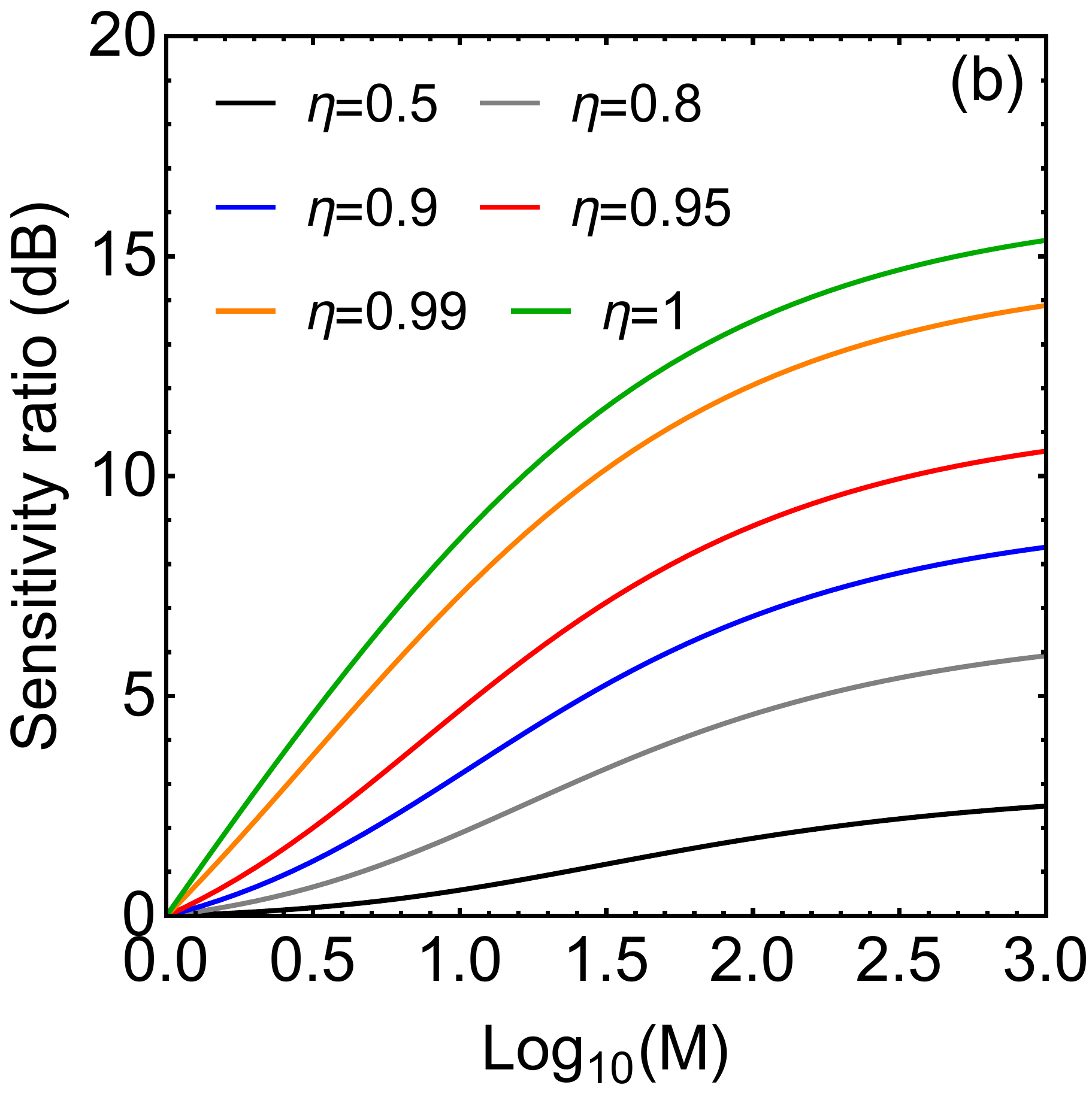}
}
\subfigure{
\centering
\includegraphics[width=0.225\textwidth]{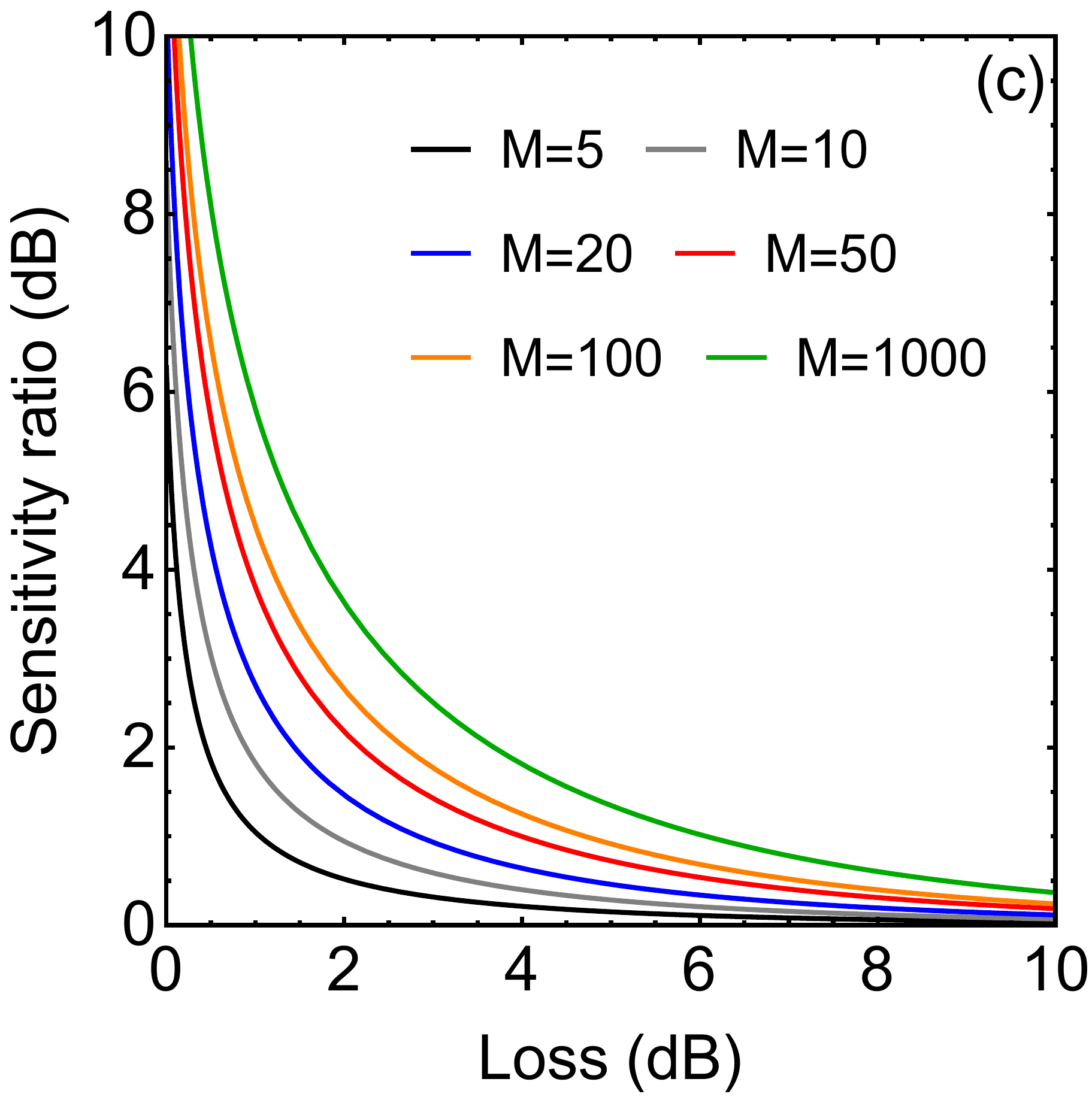}
}
\caption{ (a) Plots of the rms estimation errors, $\delta\alpha_\eta^E$ (solid curves) and $\delta\alpha_\eta^P$ (dashed curves), versus the number of sensor nodes, $M$, for various transmissivity values, from top to bottom, $\eta = 0.95, 0.99, 1$, with $n_S \equiv N_S/M$ fixed at $n_s = 1$.  (b),(c) Plots of sensitivity ratio $\left(\delta\alpha_\eta^P/\delta\alpha_\eta^E\right)^2$ in dB with $N_S = 10$. (b) From bottom to top, $\eta = 0.5, 0.8, 0.9, 0.95, 0.99, 1$. (c) From bottom to top: $M=5,10,20,50,100,1000$.
}
\label{Comparison}
\end{figure*}

\subsection{Optimum Separable State \label{OptSepState}}
To begin our derivation of the separable state  that minimizes our distributed displacement sensor's rms estimation error, it is convenient to first  constrain its input state to be a product state, $\hat{\bp}_{M,N_S} = \otimes_{m=1}^M\hat{\rho}_m$, with average photon number $N_S$.  Our entangled-state result, with $M=1$, tells us that the optimum single-mode state is the squeezed-vacuum state with average photon number $N_S$ whose real quadrature is squeezed.  It follows then that $\hat{\bp}_{M,N_S}$ must be a product of squeezed-vacuum states with squeezed real quadratures whose average photon numbers, $\{N_m\}$, satisfy $\sum_{m=1}^MN_m = N_S$.   Thus, because the $M=1$ version of $\delta\alpha^E_\eta$ is a convex function of $N_S$ for all $0<\eta\le 1$, the best product state for our sensor employs $N_m = N_S/M$ for $1\le m \le M$, leading to $\tilde{\alpha}_P \equiv \sum_{m=1}^M{\rm Re}(\hat{a}'_m)/M$ being an unbiased estimator with rms error given by 
\be
\delta \alpha_\eta^P=\frac{1}{2}\left(\frac{\eta}{M\left(\sqrt{N_S/M+1}+\sqrt{N_S/M}\right)^2}+\frac{1-\eta}{M}\right)^{1/2}.
\label{dalpha_product}
\ee
From this result we have that the optimum separable state with average photon number $N_S$ must be a $K$-fold mixture of the preceding best product states whose average photon numbers, $\{N_{S_k} : 1\le k \le K\}$, sum to $N_S$.  But the rms error in Eq.~(\ref{dalpha_product}) is a convex function of $N_S$, hence the optimum separable state for our problem must be the optimum product state specified above.  

In Appendix~\ref{AppB} we show that restricting $\hat{\bp}_{M,N_S}$ to be a \emph{Gaussian} separable state with average photon number $N_S$, and placing \emph{no} restriction on how the $\{\hat{a}'_m\}$ modes are measured to estimate $\alpha$, then the rms error is minimized by the optimum product state that we have just found, i.e., that state saturates the quantum Cram\'{e}r-Rao bound.  That said, a non-Gaussian product state could have a lower Cram\'{e}r-Rao bound for this sensing problem, but it would still have SQL scaling in the number of sensor nodes, even when $\eta=1$~\cite{refereeB2}.

\subsection{Performance Comparison} We have already seen that $\delta\alpha_\eta^E/\delta\alpha_\eta^P \simeq 1/\sqrt{M}$ when $\eta =1$ and $n_S \equiv N_S/M \gg 1$ is kept fixed.  Loss, however, quickly destroys the entangled state's Heisenberg scaling, as shown in Fig.~\ref{Comparison}(a), which plots $\log_{10}(\delta\alpha_\eta^E)$ and $\log_{10}(\delta\alpha_\eta^P)$ versus $\log_{10}(M)$ for $n_S = 1$ and various $\eta$ values.  The transmissivity required to maintain Heisenberg scaling in the number of sensor nodes when $n_S \gg 1$ is quite high:  we need $1-\eta \simeq 1/4Mn_S\ll 1$ in this case to get $\delta\alpha_\eta^E \simeq 1/2M\sqrt{2n_S}$.  Thus, absent means to realize near-lossless CVMP entanglement distribution---see below for some discussion of this point---our sensing scheme's Heisenberg scaling will be limited to local applications in which $\eta \simeq 1$ can be ensured.  Nevertheless, an appreciable performance gain can still be obtained, for moderate loss, by using CVMP entanglement, as we now show by examining performance when $N_S$, instead of $n_S$, is fixed.

Increasing $M$ with $n_S$ fixed ceases to be practical for $M \gg 1$, e.g., for Fig.~\ref{Comparison}(a)'s $M=10^4$ points the required initial squeezing is more than 40\,dB, an amount that is far beyond experimental state of the art.  So, taking $N_S = 10$, an attainable value for squeezed-vacuum generation, we plot the sensitivity ratio, $\left(\delta\alpha_\eta^P/\delta\alpha_\eta^E\right)^2$, in dB, versus $M$ for fixed $\eta$ in Fig.~\ref{Comparison}(b), and versus loss ($1/\eta$ in dB) for fixed $M$ in Fig.~\ref{Comparison}(c).  Here we see two trends:  (1) for fixed $M$ the advantage enjoyed by entangled-state operation degrades as the transmissivity decreases; and (2) for fixed $\eta$ the advantage enjoyed by entangled-state operation increases and asymptotes to a finite value as $M$ increases.  The first behavior is easily understood, i.e., it is the usual vacuum-noise degradation of nonclassical performance making the benefit of entanglement less pronounced as transmissivity decreases.  The second behavior is interesting.  For lossless ($\eta = 1$) operation with $M\rightarrow \infty$ and $N_S$ fixed, the individual states in the product-state scenario converge to vacuum states and hence $\delta\alpha_1^P \rightarrow 1/2\sqrt{M}$, while $\delta\alpha_1^E \simeq  1/4\sqrt{MN_S}$.  In this regime the $\delta\alpha_1^E/\delta\alpha_1^P = 1/2\sqrt{N_S}$ afforded by entangled-state operation matches that of lossless single-node squeezed-state operation versus lossless single-node vacuum-state operation.  Note that quadrature-displacement sensing is possible with vacuum-state inputs, because $\hat{U}(\alpha)$ converts the vacuum state to the coherent state $\ket{\alpha}$.  The final point to be drawn from Fig.~\ref{Comparison}(b) is that entangled-state operation can offer a performance gain over product-state operation for moderate loss values, e.g., at $N_S=10, M=20, \eta=0.9$, we get an 8\,dB sensitivity advantage.

\section{Applications} There are a variety of applications in which field-quadrature displacement sensing plays a central role.  
CV-QKD protocols using coherent states~\cite{grosshans2002continuous,pirandola2006continuous,pirandola2015high}, for example, rely on precise displacement operations for their security.  Our scheme can thus enable accurate joint calibration of displacement operations among multiple nodes in a quantum-secured communication network.  As seen in earlier, however, the utility of our distributed sensor for CV-QKD will be severely limited if the low transmissivity of long fiber connections cannot be mitigated.  Toward that end, continuous-variable entanglement distillation~\cite{eisert2004distillation,campbell2013continuous,ulanov2015undoing,datta2012compact} and quantum repeaters~\cite{dias2017quantum,furrer2016repeaters}, once implemented, can accomplish that mitigation.  

While awaiting developments that will permit long-distance operation of our entanglement-based displacement sensor, it has local (high-transmissivity) applications in cold-atom systems.  Quantum nondemolition detection of such systems' spin degree of freedom imprint the atoms' spin angular momentum on a light beam's field quadrature~\cite{eckert2008quantum}.  Based on this effect, measuring the temperature of a cold-atom system can be reduced to measuring an optical field's quadrature displacement~\cite{mehboudi2015thermometry}. In this scenario our scheme can reduce the rms estimation error in measuring the average temperature of a collection of locations within a cold-atom ensemble. 

\begin{figure}
\centering
\includegraphics[width=0.35\textwidth]{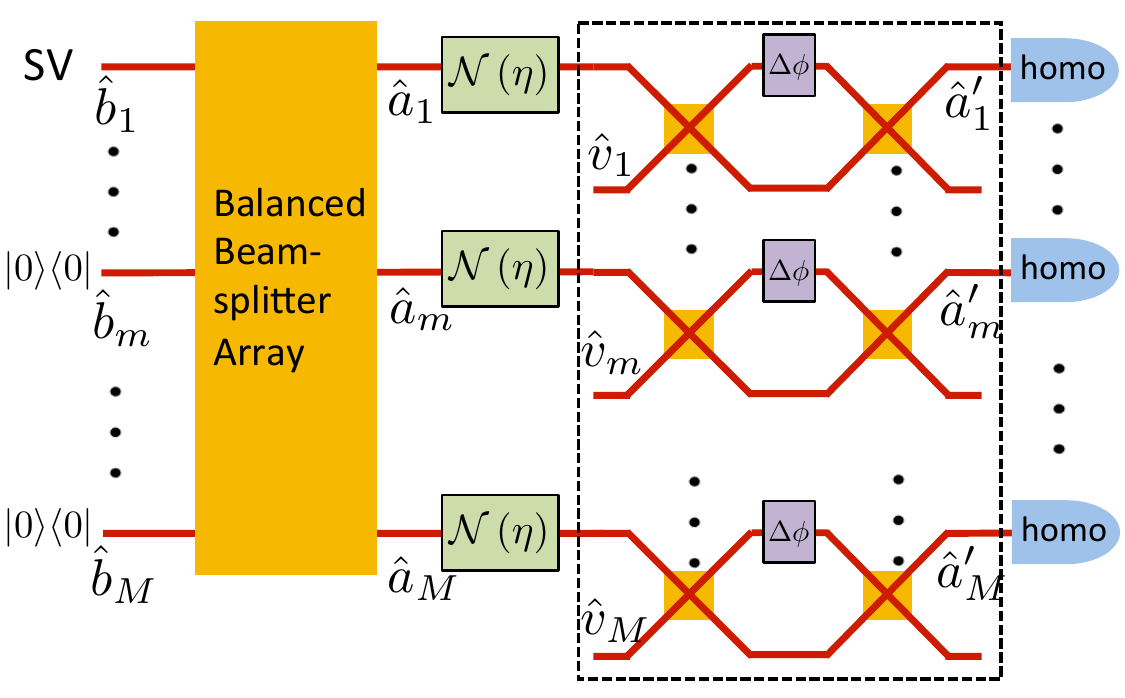}
\caption{Distributed phase-sensing interferometry.  SV: squeezed-vacuum state with mean photon number $N_S$ and squeezed noise in its imaginary quadrature. $\mathcal{N}(\eta)$:  pure-loss channel with transmissivity $0< \eta \le 1$.  
\label{int}
}
\end{figure}

Interferometric phase sensing is arguably the oldest and still widely employed optics-based sensor.  Hence successful application of our distributed displacement sensor to interferometric phase sensing would be of great significance.  Single-mode squeezed-vacuum and coherent state inputs have already been used~\cite{demkowicz2013fundamental,escher2011} to obtain quantum-enhanced performance from a phase-sensing interferometer.  Figure~\ref{int} illustrates how that arrangement can be extended to a network of Mach-Zehnder homodyne-detection interferometers that are driven by a combination of a CVMP entangled state and coherent states.  For now we assume that: all interferometers have the same $|\Delta\phi| \ll 1$ phase shift; the CVMP entangled state is obtained by passing a squeezed-vacuum state---with average photon number $N_S$ with squeezed noise in its imaginary quadrature---through the same beam splitter used in Fig.~\ref{int1}; and the $\{\hat{v}_m\}$ modes are all in their coherent state $\ket{\sqrt{N_v}}_m$.  
Then, to first order, we have that
\be 
\hat{a}_m^\prime=\sqrt{\eta}\,[(1-i\Delta \phi/2)\hat{a}_m+i\hat{v}_m\Delta \phi/2)]+\sqrt{1-\eta}\,\hat{e}_m,
\ee 
where the $\{\hat{e}_m\}$ are in their vacuum states, from which we see that $\Delta\phi$ is embedded in a field-quadrature displacement, $\alpha = i\sqrt{N_v}\,\Delta\phi/2$, of the $\{\hat{a}'_m\}$.  Thus, 
$
\widetilde{\Delta \phi}_E \equiv {2}\sum_{m=1}^M{\rm Im}(\hat{a}'_m)/{\sqrt{\eta N_v}\,M},
$
is an unbiased estimator of $\Delta \phi$ whose rms estimation error, $\delta\Delta\phi_\eta^E=2\delta\alpha_\eta^E/\sqrt{N_v}$, is 
\begin{align}
\delta\Delta\phi_\eta^E 
&= \frac{1}{\sqrt{N_v}}\sqrt{\frac{\eta}{M(\sqrt{N_S+1}+\sqrt{N_S})^2} + \frac{1-\eta}{M}}.
\end{align}
In the absence of loss, with $n_S \equiv N_S/M$ and $N_v$ fixed, this system has Heisenberg scaling in the number of interferometers.  Loss can kill this Heisenberg scaling, but for moderate loss an entanglement-based advantage---over a separable state system---still exists.  Once again, continuous-variable entanglement distillation or repeaters will be needed to make our approach suitable for widely separated interferometers.  Note that an individual interferometer from Fig.~\ref{int} was considered in~\cite{demkowicz2013fundamental,escher2011,schnabel2010quantum} as a model for improving LIGO~\cite{LIGO_nat,LIGO} by squeezed-vacuum injection (SVI).  Thus if multiple interferometers located observe correlated phase shifts---and transmission loss can be mitigated---our entanglement-based scheme can be further improve phase-sensing precision, and to do so only requires replacing product-state SVI with CVMP entangled-state injection.

\section{Conclusions} We have shown how the precision of field-quadrature displacement in a quantum network setting can be improved by use of CVMP entanglement.  Previous work~\cite{giovannetti2006} has shown that similar improvement can be obtained in a sequential manner by repeated interaction with a single system. Our scheme, however, enables all measurements to be performed simultaneously, albeit at the cost of having multiple measurement nodes, and hence is much better suited to sensing transient events.  Immediate applications of our work will likely be confined to localized sensor networks, for which high transmissivity entanglement distribution is possible.  Applications to sensor networks that span long distances will require loss-mitigation technology development to make our scheme practical. 

\begin{acknowledgements}
This research was supported by Air Force Office of Scientific Research Grant No. FA9550-14-1-0052.
QZ acknowledges support from the Claude E. Shannon Research Assistantship. 
\end{acknowledgements}

\appendix

\section{Different transmissivities and different displacements\label{AppA}}
Here we show how CVMP entanglement can be used to advantage in a modified Fig.~\ref{int1} scenario when known channel transmissivities, $\bn = (\eta_1,\eta_2,\ldots,\eta_M)$, and unknown real-valued field-quadrature displacements, $\{\alpha_m\}$, are different for each individual sensor.   In this scenario the joint state at the input to the sensor nodes' homodyne detectors is
$
\hat{\bp}_{M,N_S,\bn} = [\otimes_{m=1}^M \hat{U}_m(\alpha_m)][\otimes_{m=1}^M\mathcal{N}_m(\eta_m)](\hat{\bp}_{M,N_S})\nonumber 
[\otimes_{m=1}^M\hat{U}^\dagger_m(\alpha_m)].
$
The goal is to obtain a minimum rms error estimate of $\bar{\alpha}\equiv \sum_{m=1}^M w_m \alpha_m,$ where the weights, $\{w_m\}$, are non-negative and sum to one.  Suppose that the balanced beam splitter in Fig.~\ref{int1} is replaced with an unbalanced beam splitter that, when its nonvacuum input is a single-mode squeezed-vacuum state with average photon number $N_S$ and squeezed noise in its real quadrature, results in $\hat{b}_1 \equiv \sum_{m=1}^Mw_m\sqrt{\eta_m}\,\hat{a}_m/\bar{W}$ being in that same squeezed-vacuum state, where $\bar{W} \equiv \sqrt{\sum_{m=1}^M  w_m^2\eta_m}$.   Then, paralleling the optimality derivation presented earlier, we have that $\tilde{\alpha} \equiv \sum_{m=1}^Mw_m{\rm Re}(\hat{a}'_m)$ is an unbiased estimator of $\bar{\alpha}$ with the minimum rms error,
\be
\delta\alpha_{\bm \eta}^E = \frac{\bar{w}}{2}\left(\frac{\bar{\eta}}{(\sqrt{N_S+1}+\sqrt{N_S})^2}+1-\bar{\eta}\right)^{1/2},
\label{dalpha_extension}
\ee
under the average photon-number constraint, where $\bar{w} \equiv \sqrt{\sum_{m=1}^Mw_m^2}$ and $\bar{\eta} \equiv \sum_{m=1}^M w_m^2\eta_m/\bar{w}^2$.  

The optimal Gaussian separable-state scheme, for the scenario under consideration here, again employs a product state, but its performance
\begin{align}
&\delta \alpha_{\bm \eta}^P=\min_{\sum_{m=1}^M N_m=N_S }
\nonumber
\\
&
\left[\sum_{m=1}^Mw_m^2\left(\frac{\eta_m}{\left(\sqrt{N_m+1}+\sqrt{N_m}\right)^2}+1-\eta_m\right)/4\right]^{1/2},
\label{dalpha_product_extension}
\end{align}
cannot be found in closed form.

For a given transmissivity vector $\bn$, with $\alpha_m = \alpha$ for all $m$, we can further optimize over the $\{w_m\}$ in Eqs.~(\ref{dalpha_extension}) and (\ref{dalpha_product_extension}) to obtain minimum rms-error estimates of $\alpha$.

\section{Cram\'{e}r-Rao bound for the optimum Gaussian separable state\label{AppB}}

Here we shall obtain the quantum Cram\'{e}r-Rao (CR) lower bound on the root-mean-square (rms) estimation error $\delta\alpha_\eta^S$ of the optimum unbiased estimator for an unknown displacement $\alpha$ in the Fig.~1 setup when the joint input state to the $M$ pure-loss channels in that figure is a Gaussian separable state with total average photon number $N_S$, and no restriction is placed on the way in which the $\{\hat{a}'_m\}$ modes are measured.   From Refs.~\cite{Helstrom_1976,Holevo_1982,Yuen_1973} we have that
\be 
\delta \alpha_\eta^S\ge  \delta \alpha_\eta^{\rm CR} \equiv \sqrt{1/I_F[\hat{\rho}_{M,N_S,\eta}(\alpha)]},
\ee 
where 
\begin{eqnarray}
\lefteqn{I_F[\hat{\rho}_{M,N_S,\eta}(\alpha)]\equiv } \nonumber\\[.05in]
&&\!\!\lim_{\epsilon\to 0} 8\!\left\{1-
\sqrt{{\mathcal F}[\hat{\rho}_{M,N_S,\eta}(\alpha),\hat{\rho}_{M,N_S,\eta}(\alpha+\epsilon)]}\right\}/{\epsilon^2}
\label{Uhlmann}
\end{eqnarray}
gives the Fisher information~\cite{friis2015heisenberg} in terms of the 
Uhlmann fidelity~\cite{uhlmann1976transition},  ${\mathcal F}(\hat{\sigma}_1,\hat{\sigma}_2)\equiv \left[{\rm Tr}\!\left(\sqrt{\sqrt{\hat{\sigma}_1}\hat{\sigma}_2\sqrt{\hat{\sigma}_1}}\right)\right]^2$, between states $\hat{\sigma}_1$ and $\hat{\sigma}_2$.  The convexity of Fisher information implies that $\delta\alpha_\eta^S = \delta\alpha_\eta^P$, where $\delta\alpha_\eta^P$ is the rms error of the optimum unbiased estimator of $\alpha$ for a Gaussian \emph{product} state, under the same average photon-number constraint.  

Let $\hat{\bp}_{M,N_S}=\otimes_{m=1}^M \hat{\rho}_{N_{m}}$ be the Gaussian product-state input to the pure-loss channels in  Fig.~1, where $\hat{\rho}_{N_{m}}$, the Gaussian state sent to the $m$th sensor node, has average photon number $N_m$ and $\sum_{m=1}^M N_m = N_S$.  The joint state at the inputs to the sensors nodes' quantum measurements is thus $\hat{\bp}_{M,N_S,\eta}(\alpha)=\otimes_{m=1}^M \hat{\rho}_{N_m,\eta}(\alpha)
$, with $\hat{\rho}_{N_m,\eta}(\alpha)\equiv \hat{U}(\alpha)[\mathcal{N}(\eta)(\hat{\rho}_{N_{m}})]\hat{U}^\dagger(\alpha)$, which is also a Gaussian product state.

For product states we have that $I_F[\hat{\bp}_{M,N_S,\eta}(\alpha)]=\sum_{m=1}^M I_F[\hat{\rho}_{N_m,\eta}(\alpha)]$, so the optimum product state's rms error satisfies
\be
\delta \alpha_\eta^P\ge \delta\alpha_\eta^{\rm CR} = \min_{\hat{\bp}_{M,N_S}} \sqrt{1\big/\sum_{m=1}^M I_F[\hat{\rho}_{N_m,\eta}(\alpha)]}.
\ee
To evaluate this minimum we will first find $\max_{\hat{\rho}_{N_m}} I_F[(\hat{\rho}_{N_m,\eta}(\alpha)]$, when $\hat{\rho}_{N_m}$ is a single-mode Gaussian state with average photon number $N_m$.   

The single-mode Gaussian state $\hat{\rho}_{N_m}$ is completely characterized~\cite{Weedbrook2012} by its quadratures' mean vector ${\boldsymbol a}_m$ and covariance matrix ${\bf V}_m$, where we take those quadratures to be ${\rm Re}(\hat{a}_m)$ and ${\rm Im}(\hat{a}_m)$.  Then, writing $\hat{\rho}_{N_m}$ as the Gaussian state $\hat{\rho}_G({\boldsymbol a}_m,{\bf V}_m)$, we get $\hat{\rho}_G(\sqrt{\eta}\,{\boldsymbol a}_m + {\boldsymbol \alpha}, \eta{\bf V}_m + (1-\eta){\bf I}/4)$ for the Gaussian state $\hat{\rho}_{N_m,\eta}(\alpha)$, where ${\boldsymbol \alpha} \equiv [\alpha,0]$ and ${\bf I}$ is the $2\times 2$ identity matrix, and we get
$\hat{\rho}_G(\sqrt{\eta}\,{\boldsymbol a}_m + {\boldsymbol \alpha} + {\boldsymbol \epsilon}, \eta{\bf V}_m + (1-\eta){\bf I}/4)$ for the Gaussian state $\hat{\rho}_{N_m,\eta}(\alpha+ \epsilon)$, where ${\boldsymbol \epsilon} \equiv [\epsilon,0]$.  The quadrature covariance matrix of an arbitrary $\hat{\rho}_G({\boldsymbol a}_m,{\bf V}_m)$ can always be written in the form ${\bf V}_m = {\bf R}_\theta {\bf V}_{\rm diag}{\bf R}_\theta^T$, where ${\bf V}_{\rm diag} =  {\rm Diag}[(2n_m+1)e^{-r_m}/4,(2n_m+1)e^{r_m}/4]$ with $r_m\ge 0$, $n_m\ge 0$, and 
\be
{\bf R}_\theta = \left[\begin{array}{cc}
\cos(\theta) & \sin(\theta) \\
-\sin(\theta) & \cos(\theta) \end{array}
\right].
\ee
With this ${\bf V}_m$ representation we can use Ref.~\cite{scutaru1998fidelity} to evaluate the Uhlmann fidelity between $\hat{\rho}_G(\sqrt{\eta}\,{\boldsymbol a}_m + {\boldsymbol \alpha},\eta {\bf V}_m + (1-\eta){\bf I}/4)$ and $\hat{\rho}_G(\sqrt{\eta}\,{\boldsymbol a}_m + {\boldsymbol \alpha} + {\boldsymbol \epsilon}, \eta{\bf V}_m + (1-\eta){\bf I}/4)$.  Using the result of that evaluation in Eq.~(\ref{Uhlmann}) gives us
\begin{widetext}
\begin{equation}
I_F[\hat{\rho}_{N_m,\eta}(\alpha)] = 
\frac{4\{e^{r_m}(1-\eta) + (2n_m+1)\eta[e^{2r_m}\cos^2(\theta) + \sin^2(\theta)]\}}{(e^{r_m}(1-\eta)+(2n_m+1)\eta)[(2n_m+1)\eta e^{r_m} + 1-\eta]}.
\end{equation}
\end{widetext}
This expression's maximum over $\theta$ and $n_m$ occurs when $\theta = n_m = 0$, in which case we get $\max_{\theta,n_m}I_F[\hat{\rho}_{N_m,\eta}(\alpha)]=4 /(\eta e^{-r_m}+ 1-\eta)$ with $N_m = {\boldsymbol a}_m^T{\boldsymbol a}_m + [\cosh(r_m) - 1]/2$.  From this result it is clear that ${\boldsymbol a}_m = {\bf 0}$ is optimum, and we find that
\begin{eqnarray}
\lefteqn{\max_{\hat{\rho}_{N_m}}I_F[\hat{\rho}_{N_m,\eta}(\alpha)]
= }\nonumber \\[.05in]
&& \left(\frac{\eta}{4(\sqrt{N_m+1} + \sqrt{N}_m)^2} + \frac{1-\eta}{4}\right)^{-1}.
\end{eqnarray}

At this point we have that
\begin{eqnarray}
\lefteqn{\delta\alpha_\eta^S = \delta\alpha_\eta^P\ge \delta\alpha_\eta^{\rm CR} = \min_{\sum_{m=1}^MN_m = N_S} } \nonumber \\[.05in]
&&  \frac{1}{2}\left[\sqrt{\sum_{m=1}^M\left(\frac{\eta}{(\sqrt{N_m+1} + \sqrt{N}_m)^2} + 1-\eta\right)^{-1}}\right]^{-1}.
\end{eqnarray}
Because $\max_{\hat{\rho}_{N_m}} I_F\left(\hat{\rho}_{N_m,\eta}\left(\alpha\right)\right)$ is a concave function of $N_m$, the preceding minimum is achieved by $N_m = N_S/M$ for $1\le m \le M$, hence we have the quantum CR bound for Gaussian separable states:  
\begin{eqnarray}
\lefteqn{\delta\alpha_\eta^S = \delta\alpha_\eta^P\ge \delta\alpha_\eta^{\rm CR} = \frac{1}{2}} \nonumber \\[.05in]
&&\!\!\!\times\! \left(\frac{\eta}{M\left(\sqrt{N_S/M+1}+\sqrt{N_S/M}\right)^2}+\frac{1-\eta}{M}\right)^{1/2}\!\!\!\!. 
\label{dalpha_product_supp}
\end{eqnarray}
We showed in Sec.~\ref{OptSepState} that this CR bound performance is achieved by modal homodyne detection using the estimator $\tilde{\alpha}_P = \sum_{m=1}^M{\rm Re}(\hat{a}'_m)/M$ when $\hat{\bp}_{N_S,M}$ is an $M$-fold tensor product of squeezed-vacuum states each with average photon number $N_S/M$ and squeezed noise in their real quadratures, i.e., an $M$-fold tensor product of zero-mean Gaussian states with covariance matrix ${\bf V} = {\rm Diag}[e^{-r}/4, e^r/4]$ and $\cosh(r) = 2N_S/M + 1$, as found above.


\begin{thebibliography}{36}
\bibitem{footnote3}Homodyne-detection efficiencies are typically quite high, and any inefficiencies therein can be included in the channel transmissivities with minor revisions to our derivations.
\end{thebibliography}
\end{document}